\documentclass[aps,prl,amsmath,amssymb,reprint]{revtex4-2}
\usepackage{amsmath, amsthm, amssymb}
\usepackage{hyperref}
\usepackage{xfrac}
\usepackage{mathtools}
\usepackage{verbatim}
\usepackage{epsfig,graphicx}
\usepackage[english]{babel}

\makeatletter
\newcommand\gsl{\ifmmode\textsl{g}\else g\fi}
\makeatother

\begin{document}

\title{Heat spots as structural elements of synoptic turbulence model}

\author{V.~P.~Goncharov}
\email{v.goncharov@rambler.ru}
\affiliation{A. M. Obukhov Institute of Atmospheric Physics RAS, 109017 Moscow, Russia}

\date{\today}

\begin{abstract}
The large-scale dynamics of heat spots in a thin layer of incompressible rotating fluid under the action of Coriolis and gravity is considered to obtain a simple model of synoptic turbulence. The derivation of equations describing the evolution of the spots is based on the use of the variational principle, perturbation theory, and the assumption of geostrophic balance. Exact solutions with piecewise constant contour curvature are found. The simplest of them look like circular spots that move with constant velocity depending on their radius. More complex solutions, such as ``loop solitons'', are shown to have fractal structure and are constructed by the gluing method. Heat spots can act as structural elements of turbulence. In particular, we show that the spectral energy density for the velocity field of a random ensemble of heat spots has the power asymptotics $E\propto k^{\alpha}$ with exponents $\alpha=2$ at $k\bar{R}<1$, and $\alpha=-1$ at $k\bar{R}>1$, where $\bar{R}$ is the average (over ensemble) radius of spot.
\end{abstract}

\maketitle

{\it Introduction.}---Large-scale atmospheric dynamics learns of motion patterns on scales of several hundred kilometers or more~\cite{hh12}. On such scales, referred to as synoptic, the atmosphere can be viewed as a thin layer of incompressible horizontally inhomogeneous fluid rotating in the gravity field. As shown in Ref.~\cite{g21}, in the case when fluid motions are sufficiently slow and localized in thin jets generated by sharp buoyancy gradients around heat spots, one can reformulate theory so that the description of such spots in terms of $2D$-contour dynamics becomes possible.

The main aim of this paper is to explain some spectral features inherent to synoptic turbulence by relying on simplified model assumptions. In contrast to the conventional theory of turbulence~\cite{my07}, the approach utilized here assumes that the base elements of turbulence are not random field fluctuations but heat spots -- exact solutions of a nonlinear problem with random free parameters.

Since heat spots arise from complex nonlinear interactions in the dynamical system, they initiate motions spanning over a wide range of spatio-temporal scales, the coupling of which can lead to chaos and loss of the dynamical predictability. For this reason, procedures of discarding fast small-scale processes from a complete description to obtain a large-scale model always raise many questions. One of them, the most important, is whether the assumption of a weak influence of small-scale chaos on large-scale motions, implicitly implied in this paper, is valid or not. At least the conclusion in Ref.~\cite{abr12} that chaos in ``slow'' variables can be suppressed by the rapidly mixing ``fast'' variables, up to the extent that the behavior of the slow variables becomes predictable, answers this question affirmatively. And this inspires a certain optimism.

{\it Contour dynamics and integrals of motion.}---For the planar model of heat spots, the equations of motion are derived in the inertial-free approximation from the Lagrangian
\begin{equation}
\mathcal{L}=-\frac{H}{2}\int\left(f\hat{x}_{t}\hat{y}-
f\hat{y}_{t}\hat{x}-\gsl^{\prime}\frac{H}{J}b\right)dsdn.\label{eq:1}
\end{equation}
Here the average layer depth $H$ and the Coriolis parameter $f$ are constants, the fluid particles coordinates $\hat{x}$, $\hat{y}$ depend on both labeling coordinates $(s,n)$ and time $t$, and the relative buoyancy distribution $b$ is a function of the labeling coordinates only. The quantity $J$, termed the Jacobian, is given by the relation
\begin{equation}
J=\hat{x}_{s}\hat{y}_{n}-\hat{x}_{n}\hat{y}_{s}.\label{eq:2}
\end{equation}

To formulate the problem in a dimensionless form, the quantity $T=f^{-1}$ and the Rossby deformation radius $L=(\gsl^{\prime}H)^{1/2}/f$, where $\gsl^{\prime}$ is the reduced-gravity constant, are convenient to take as scales of time and horizontal length. Then, as shown in Ref.~\cite{g21}, after a non-dimensionalization by the rules
\begin{equation}
\left(\hat{x},\hat{y},s,n\right)\rightarrow
L\cdot\left(\hat{x},\hat{y},s,n\right),\quad t\rightarrow Tt,\label{eq:3}
\end{equation}
in the thin jet approximation, the Lagrangian (\ref{eq:1}) reduces to the form
\begin{equation}
\mathcal{L}=\int\Biggl(Y_{t}X-X_{t}Y+\gamma\frac{Y_{s}X_{ss}-
X_{s}Y_{ss}}{\left(X_{s}^{2}+Y_{s}^{2}\right)^{2}}\Biggr)ds.\label{eq:4}
\end{equation}
The variables $X(s,t)$, $Y(s,t)$ included in this expression describe trajectories along which thin jets localized on buoyancy jumps circulate about heat spots. The occurring mechanism of such jets is the geostrophic wind effect~\cite{hh12}. The time-independent function $\gamma(s)$ characterizes the transversal buoyancy gradient at these trajectories. In the case of a single isolated heat spot, the coordinates $X(s,t)$ and $Y(s,t)$ describe the time evolution of a closed contour.

The contour dynamic equations formulated for the variables $X(s,t)$, $Y(s,t)$ follow directly from the variational action principle. In that simple case, when $\gamma=1$, they are written in the form
\begin{equation}
X_{t}=\partial_{s}\frac{X_{ss}}{\left(X_{s}^{2}+Y_{s}^{2}\right)^{2}},\quad
Y_{t}=\partial_{s}\frac{Y_{ss}}{\left(X_{s}^{2}+Y_{s}^{2}\right)^{2}},\label{eq:5}
\end{equation}
and, being Hamiltonian, can be presented as
\begin{gather}
X_{t}=-\frac{\delta\mathcal{H}}{\delta Y},\quad
Y_{t}=\frac{\delta\mathcal{H}}{\delta X},\label{eq:6}\\
\mathcal{H}=\frac{1}{2}\int\frac{Y_{ss}X_{s}-X_{ss}Y_{s}}
{\left(X_{s}^{2}+Y_{s}^{2}\right)^{2}}ds.\label{eq:7}
\end{gather}
Aside from the Hamiltonian $\mathcal{H}$, which represents the total energy, Eqs.~(\ref{eq:5}), (\ref{eq:6}) conserve still several closed-loop integrals.
\begin{gather}
I_{X}=\int Xds,\quad I_{Y}=\int Yds,\label{eq:8}\\
P=\frac{1}{2}\int\left(X^{2}+Y^{2}\right)ds,\quad
A=\frac{1}{2}\int\left(X_{s}Y-Y_{s}X\right)ds.\label{eq:9}
\end{gather}
Among them, the last two play the role of the moment of inertia and the spot area, respectively.

{\it Round spot solutions.}---The main difficulty in studying heat spot evolution bears a mathematical nature and relates to the closedness of their contour. However, the same feature endows them with the property of corpuscularity that allows them to pretend to be structural elements in many physical processes, like turbulence or heat transport.

One of these solutions we can derive from Eqs.~(\ref{eq:5}) by rewriting them in a more convenient form. The corresponding description version follows from a variable replacement, so $X$, $Y$, and $s$ are regarded next as functions of a new independent variable $\sigma$:
\begin{equation}
X=X\left(\sigma,t\right),\quad Y=Y\left(\sigma,t\right),\quad
s=s\left(\sigma,t\right).\label{eq:10}
\end{equation}
The freedom in the choice of the variable $s$s leaves us many possibilities to simplify and reformulate the original problem.

In particular, taking the arc length as such a variable, we get the constraint
\begin{equation}
\quad X_{\sigma}^{2}+Y_{\sigma}^{2}=1.\label{eq:11}
\end{equation}
Then, the rules
\begin{equation}
\partial_{t}\rightarrow\partial_{t}-\frac{s_{t}}{s_{\sigma}}\partial_{\sigma},\quad
\partial_{s}\rightarrow\frac{1}{s_{\sigma}}\partial_{\sigma},\label{eq:12}
\end{equation}
ensuring by transformations (\ref{eq:10}), allow us to reduce Eqs.~(\ref{eq:5}) into the following form
\begin{gather}
X_{t}s_{\sigma}-s_{t}X_{\sigma}=\partial_{\sigma}\left(s_{\sigma}\left(s_{\sigma}
X_{\sigma\sigma}-s_{\sigma\sigma}X_{\sigma}\right)\right),\label{eq:13}\\
Y_{t}s_{\sigma}-s_{t}Y_{\sigma}=\partial_{\sigma}\left(s_{\sigma}\left(s_{\sigma}
Y_{\sigma\sigma}-s_{\sigma\sigma}Y_{\sigma}\right)\right).\label{eq:14}
\end{gather}
Note that using the arc length as an independent variable changes the type of nonlinearity of motion equations, allowing us to eliminate it in the denominators. This technique is effective not only for finding partial solutions but sometimes the problem, with its help, becomes completely integrable.

The simplest solution of Eqs.~(\ref{eq:13}), (\ref{eq:14}) looks like a heat spot of round shape moving with a constant velocity. One can find exactly such a solution by assuming that
\begin{equation}
s=aX+bY,\label{eq:15}
\end{equation}
and
\begin{equation}
X=X_{0}+\kappa^{-1}\sin\left(\kappa\sigma\right),\quad
Y=Y_{0}-\kappa^{-1}\cos\left(\kappa\sigma\right).\label{eq:16}
\end{equation}
Here $a$, $b$ are some constants, $\kappa=1/R$ is the constant curvature, $R$ is the spot radius, and $X_{0}(t)$, $Y_{0}(t)$ are its time-dependent coordinates.

The substitution of (\ref{eq:15}), (\ref{eq:16}) into (\ref{eq:13}) and (\ref{eq:14}) leads to constraints
\begin{equation}
a=R^{2}\dot{X}_{0},\quad b=R^{2}\dot{Y}_{0},\label{eq:17}
\end{equation}
where dots over quantities denote time derivatives. From this, we can conclude that the round-shaped spots move with constant velocities determined only by their radius and the parameters $a$, $b$.

{\it Stability of round heat spots.}---An important physical property of any nonlinear solution is its stability. To analyse this effect for round heat spots, without any loss of generality, let us assume that such a spot has the unit radius and moves along the $X$-axis with the unit speed. Doing so and returning to the independent variable $s$, in a comoving frame of reference, we come to the perturbed solution
\begin{equation}
X=t+s+\beta,\quad Y=\sqrt{1-s^{2}}+\alpha,\quad|s|\leq1,\label{eq:18}
\end{equation}
where the variables $\alpha(s,t)$, $\beta(s,t)$ describe fluctuations. If perturbations are sufficiently small, as follows from both Eqs.~(\ref{eq:5}) and the motion invariants (\ref{eq:7})--(\ref{eq:9}), they obey the equations
\begin{gather}
\alpha_{\tau}-\alpha_{s}=\partial_{s}\left[\partial_{s}\left(1-s^{2}\right)^{2}
\alpha_{s}+4\left(1-s^{2}\right)^{3/2}\beta_{s}\right],\label{eq:19}\\
\beta_{\tau}-\beta_{s}=\partial_{s}^{2}\left(1-s^{2}\right)^{2}\beta_{s},\label{eq:20}
\end{gather}
and, on the other hand, must satisfy the constraints
\begin{gather}
\quad\int\beta ds=0,\quad\int\beta_{s}\sqrt{1-s^{2}}ds=0,\label{eq:21}\\
\int\alpha ds=0,\quad\int\left(\alpha\sqrt{1-s^{2}}+s\beta\right)ds=0.\label{eq:22}
\end{gather}

The subsequent numerical stability analysis based on expansions of perturbations into finite series of Legendre polynomials and iterative calculations gives a trivial result $\alpha=0$, $\beta=0$. This indicates the stability of the moving round-shaped spots, at least in the linear approximation.

\begin{figure}[t!]
\centering{
\begin{minipage}[h]{\linewidth}
\includegraphics[width=\columnwidth]{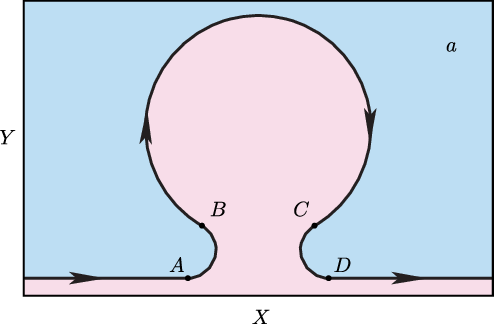}
\end{minipage}\\
\vfill\vspace{0.2cm}
\begin{minipage}[h]{\linewidth}
\includegraphics[width=\columnwidth]{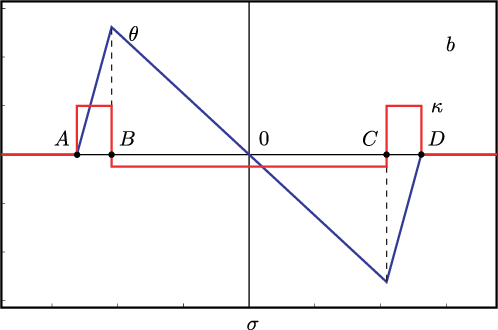}
\end{minipage}]}
\caption{(a) The illustration of a fractal-loop solution. The rose color denotes a warm fluid, while the blue color indicates a cold fluid. The arrows show the direction of circulation along the jet trajectory. (b) The curvature $\kappa$ (in red) and slope angle $\theta$ (in blue) are depicted as functions of the arc length $\sigma$.}\label{fig1}
\end{figure}

{\it Fractal solutions with piecewise-constant curvature.}---More general solutions can be constructed from arcs with piecewise-constant curvature by connecting them, one by one, at inflection points. Fig.~\ref{fig1} illustrates an example of such a solution.
It describes the trajectory that runs along buoyancy jumps and includes a fractal loop. The whole structure moves as a unit without deformations at a constant speed. Since the buoyancy jumps are responsible for the thin jets, and those, in turn, generate circulation, this quantity is possible to evaluate as the integral
\begin{equation}
C=\int\left(X_{t}dX+Y_{t}dY\right)=
-\gamma\int\frac{X_{ss}^{2}+Y_{ss}^{2}}{\left(X_{s}^{2}+
Y_{s}^{2}\right)^{2}}ds.\label{eq:23}
\end{equation}
Whence it follows that the circulation $C$ is opposite in sign to the buoyancy gradient $\gamma$.

Since the fractal solutions are lines with piecewise-constant curvature, they are not fixed wholly in the plane by their free parameters -- propagation speed and the radii of the arcs making them up. There exists one more source of arbitrariness -- the gluing procedure. Such a parametric uncertainty opens the possibility of constructing irregular fractal solutions whose all free parameters are random quantities. One application area of fractal solutions is the development of turbulence models~\cite{ph58,saf71,kp73,kuz04,knnr07,gp14} by using the randomization of free parameters.

{\it Structural turbulence and its spectra.}---Let us consider turbulence in the frame of the given model as a collective effect created by a statistical ensemble of temperature spots. It is worth noting that a similar approach, but in  different contexts, a similar approach had been applied in Refs.~\cite{kuz04,knnr07,gp14}. To exclude the influence of all other factors except statistics, we restrict ourselves to some simple model assumptions by neglecting heat spot interactions between themselves.

Since, in the Eulerian coordinates $\mathbf{r}=(x,y)$, the fields of velocity $\mathbf{u}(\mathbf{r},t)$ and buoyancy $b(\mathbf{r},t)$ are linked by the condition of geostrophic balance, in dimensionless form, they obey the relation
\begin{equation}
\mathbf{u}=\frac{1}{2}{\boldsymbol{\nabla}^\perp}b,\label{eq:24}
\end{equation}
where the symbol $\perp$ denotes the $2D$-vector rotation by 90 degrees anticlockwise. So, the operation ${\boldsymbol{\nabla}^\perp}=\left(-\partial_{y},\partial_{x}\right)$ is none other than the skew-symmetric gradient.

Within the above model, each heat spot occupies a finite area where its buoyancy takes a constant value. In particular, if it equals $\gamma$, its space distribution can be given analytically by Cauchy's integral
\begin{equation}
b=\frac{i\gamma}{2\pi}\oint\frac{Z_{s}ds}{z-Z}.\label{eq:25}
\end{equation}
This integral is taken in the complex plane $z=x+iy$ over the closed contour $Z=X+iY$ and, after differentiation with the operator ${\boldsymbol{\nabla}^\perp}$, allows to get the velocity field
\begin{equation}
\mathbf{u}=-\frac{\gamma}{2}\oint\mathbf{Z}_{s}\delta
\left(\mathbf{r-Z}\right)ds,\quad\mathbf{Z}=\left(X,Y\right).\label{eq:26}
\end{equation}
Dirac's $\delta$-function under the integral sign says that fluid motions are localized at spot boundaries $\mathbf{Z}(s,t)$ and, hence, have a jet-like pattern.

In the Fourier $\mathbf{k}$-space, from Eq.~(\ref{eq:26}) it follows that
\begin{equation}
\mathbf{u}_{\mathbf{k}}=-\frac{\gamma}{4\pi}\oint\mathbf{Z}_{s}
e^{-i\mathbf{k\cdot Z}}ds.\label{eq:27}
\end{equation}
Thus, for a round spot moving with velocity $\mathbf{c}$, it can be found that
\begin{equation}
\mathbf{u}_{\mathbf{k}}=i\frac{\gamma R}{2k}\mathbf{k}^{\perp}
e^{-i\mathbf{k\cdot}\left(\mathbf{Z}^{0}+\mathbf{c}t\right)}J_{1}(kR),\label{eq:28}
\end{equation}
where $\mathbf{Z}^{0}$ are initial coordinates of the center of the spot, $R$ is its radius, $J_{1}$ is the Bessel function of the first kind.

If the round spots are distributed freely enough without overlapping, the spectrum for the whole ensemble of such spots can be obtained by summing the spectra of individual spots. Thus, according to the superposition principle, we obtain
\begin{equation}
\mathbf{u}_{\mathbf{k}}=i\frac{\mathbf{k}^{\perp}}{2k}\sum_{\alpha}\gamma_{\alpha}R_{\alpha}
e^{-i\mathbf{k\cdot}\left(\mathbf{Z}^{0}_{\alpha}+
\mathbf{c}_{\alpha}t\right)}J_{1}(kR_{\alpha}).\label{eq:29}
\end{equation}
Free parameters of spots listed here by index $\alpha$ will now play the role of independent random variables.

A calculation of the energy density spectrum $E(\mathbf{k})$ proceeds using the standard procedure \begin{equation}
E=\frac{1}{2}\langle|\mathbf{u}_{\mathbf{k}}|^{2}\rangle,\label{eq:30}
\end{equation}
where angle brackets imply averaging. Since the random variables are independent, the averaging procedure factorizes, i.e., it becomes a product of three averages
\begin{equation}
E=\frac{1}{8}n\langle\gamma^{2}\rangle\varepsilon(\mathbf{k}),\quad
\varepsilon(\mathbf{k})=\Big\langle R^{2}J_{1}^{2}(kR)\Big\rangle.\label{eq:31}
\end{equation}
Here, the first constant $n$ denoting spot density on the plane arises from averaging over the coordinates $\mathbf{Z}^{0}_{\alpha}$ provided their distribution uniformity. The second quantity $\langle\gamma^{2}\rangle$ implies the mean square of buoyancy. The answer of how the energy density spectrum depends on the wavevector $\mathbf{k}$ needs to do one more averaging over spot radii $R$.

\begin{figure}[t!]
\includegraphics[width=\columnwidth]{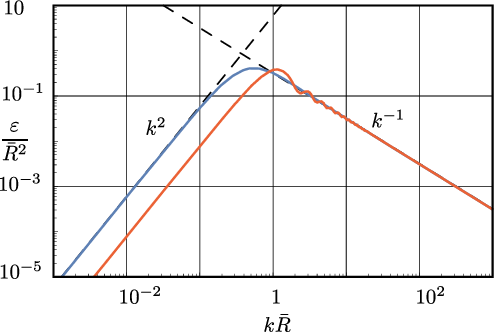}
\caption{The energy spectrum of turbulence resulting from the generation of round spots. The dashed lines on the log-log plot show power asymptotes. The red plot indicates the uniform distribution, while the blue plot corresponds to the exponential distribution (\ref{eq:33}).}\label{fig4}
\end{figure}

As follows from the maximum-entropy principle~\cite{j157,j257,j03,ct06}, an appropriate distribution function $f(R)$ looks like
\begin{equation}
f=\bar{R}^{-1}e^{-R/\bar{R}},\label{eq:33}
\end{equation}
where $\bar{R}$ is a finite mean radius of spots. Averaging with this distribution gives
\begin{equation}
\varepsilon(\mathbf{k})=2\frac{\bar{R}^{2}}{\pi}\frac{1-4\varrho^2}{1+4 \varrho^2}
\left(\frac{K\left(-4\varrho^2\right)}{1-4\varrho^2}-\frac{E\left(-4\varrho^2\right)}
{1+4 \varrho^2}\right),\label{eq:34}
\end{equation}
where $K$ and $E$ are elliptic integrals of the first and second kind, respectively. The spectral $k$-dependence realises through the parameter $\varrho=k\bar{R}$ and is shown in Fig.~\ref{fig4}.

{\it Discussion and perspectives.}---The model of large-scale turbulence considered here includes only those approximations and idealizations that, ignoring the rest, assume the influence of only two factors - rotation and gravity. That is why the results we got have a qualitative character. The main aim of the model is to show that not only external factors (sources) but also the nonlinearity of the medium can be responsible for the formation of local maxima and power laws in the spectra of atmospheric turbulence. In fluid dynamics, as well known, nonlinearity leads to the formation of spatially localized structures that remain stable about various kinds of perturbations. By conserving nontrivial integrals of motion, these structures comply with internal, hidden symmetries of the nonlinear medium and manifest themselves against the background of small-scale turbulence as domains of highly ordered (coherent) large-scale motions.

Figure~\ref{fig4} suggests that large-scale (synoptic) turbulence due to heat spots features a well-defined maximum and two power-law asymptotes: $k^{2}$ to the left and $k^{-1}$ to the right of it. It is noteworthy that altering the distribution functions does not impact the power laws. It only slightly shifts the position of the spectral maximum along the $k$-axis. The explanation for such universality lies in that, even before averaging, the Fourier spectrum of the velocity field of a round heat spot already has power-law asymptotics.

The spectral slope with $k^{-1}$ aligns with the results of Refs.~\cite{zjbb17,st22}, where global model simulations were applied to study atmospheric energy spectra. As found, this slope belongs to the long-wave part of the kinetic energy spectrum and, spanning the range $1\leq n\leq7$ of non-dimensional zonal wavenumbers $n$, transits to the spectrum with the slope of $-3$, when $n$ becomes greater $7$.

It is evident that other stable solutions to the problem (\ref{eq:5}) can also serve as structural elements of turbulence. The choice of circular temperature spots in this role was only due to the simplicity of these solutions. Of course, extending the statistical ensemble due to fractal and other solutions can contribute some spectral refinements and corrections in the spectral behavior of turbulence. This step, however, does not touch the symmetries (translational invariance and isotropy), which prevent the applicability of the plane model on planetary scales. From this viewpoint, a more promising approach would be to study the structural turbulence not on a rotating plane but on a rotating sphere where solutions like heat spots also exist, although in a different shape~\cite{g23}.

\acknowledgments
This work was supported by the Russian Science Foundation No. 23-17-00273.

\end{document}